# Metamaterial analogue of Ising model

Longqing Cong,[1,2] Vassili Savinov,[3] Yogesh Kumar Srivastava,[1,2] Song Han,[1,2] and

Ranjan Singh[1,2,*]

[1]*Division of Physics and Applied Physics, School of Physical and Mathematical Sciences, Nanyang Technological University, Singapore 637371, Singapore*
[2]*Centre for Disruptive Photonic Technologies, The Photonics Institute, School of Physical and Mathematical Sciences, Nanyang Technological University, Singapore 637371, Singapore*
[3]*Optoelectronics Research Centre and Centre for Photonic Metamaterials, University of Southampton, Southampton, SO17 1BJ, United Kingdom*

[*]*E-mail:* ranjans@ntu.edu.sg

**Abstract**

The interaction between microscopic particles has always been a fascinating and intriguing area of science. Direct interrogation of such interactions is often difficult or impossible. Structured electromagnetic systems offer a rich toolkit for mimicking and reproducing the key dynamics that governs the microscopic interactions, and thus provide an avenue to explore and interpret the microscopic phenomena. In particular, metamaterials offer the freedom to artificially tailor light-matter coupling and to control the interaction between unit cells in the metamaterial array. Here we demonstrate a terahertz metamaterial that mimics spin-related interactions of microscopic particles in a 2D lattice via complex electromagnetic multipole interactions within the metamaterial array. Fano resonances featured by distinct mode properties due to strong nearest-neighbor interactions are discussed that draw parallels with the 2D Ising model. Interestingly, a hyperfine Fano splitting spectrum is observed by manipulating the 2D interactions without applying external magnetic or electric fields, which provides a passive multispectral platform for applications in super-resolution imaging, biosensing, and selective thermal emission. The dynamic approach to reproduce the static interaction between microscopic particles would enable more profound significance in exploring the unknown physical world by the macroscopic analogues.

Interaction between large aggregates of particles lies at the heart of our understanding of complex macroscopic behaviors exhibited by solids, gases, and fluids near phase transitions.[1] Models of particle interactions provide a framework for study and classification of critical phenomena, and are thus important in modelling a wide variety of scenarios ranging from solid state physics to high-energy physics, biology of complex systems, and even economics.[2] One of the simplest models that captures a significant share of dynamics of this kind is the two-dimensional (2D) Ising model[3]: a 2D lattice of particles which have a 'spin', i.e. a magnetic dipole moment. The energy of this system depends on the orientation of different spins relative to each other as well as relative to an externally applied magnetic field. The analytical description of the basic 2D Ising model has been available since 1944,[4] but the model remains of interest both in the context of understanding properties of Ising models in higher dimensions, as well as describing 2D Ising-like systems with complex interactions that are not captured by the analytical description. In this work we show how metamaterials, a class of man-made electromagnetic media, can be used to mimic and therefore experimentally explore the properties of 2D Ising models.

Metamaterials are artificial media created by patterning unit cells on the scale smaller than the target wavelength of the electromagnetic excitation.[5,6] Here we investigate metamaterials composed of terahertz asymmetric split ring resonators (TASR) housed on a low-loss dielectric substrate, as shown in Fig. 1a. Each TASR is a metallic square with split gaps at the top and bottom sides. The top split is displaced horizontally from the vertical line of symmetry of TASR with a distance *d*, therefore, the TASR is split into two metallic segments of different length. Due to asymmetry in segment length, incident light at normal direction polarized along the metallic segments (*y*-axis in Fig. 1a) can excite anti-symmetric current oscillations in the two segments. This is known as a trapped mode resonance[7] as well as Fano resonance.[8-10] At the Fano resonance, the anti-symmetric current oscillations minimize the energy lost to free space, resulting in a narrow resonance mode.[7]

All the discussion in this work will center on responses of TASR metamaterials at the Fano resonance, and in particular on interactions between the individual TASR. We therefore introduce a convenient model of a single TASR operating at the trapped mode, as shown in Fig. 1b, which represents it as two electric dipoles and one magnetic dipole. As discussed above, the Fano resonance corresponds to opposing current oscillations in the two metallic segments of each TASR, which would give rise to an out-of-plane magnetic dipole ($M_z$). In an asymmetric resonator, the Fano mode also corresponds to electric dipoles along *y*-axis ($P_y$) as well as *x*-axis ($P_x$), which arises due to lack of mirror symmetry in vertical axis. An important property of TASR metamaterials, at the Fano resonance, is that strong magnetic-dipole-mediated interactions between the neighboring TASRs lead to emergence of the so-called collective mode or cooperative resonance,[11-14] where the whole metamaterial behaves as a single planar cavity (see Fig. 1a). In this work, we will demonstrate the direct analogy between the interactions of dynamic dipoles in TASR metamaterials and the interactions of static dipoles, which are important in Ising model, as well as in shaping magnetic and electric ordering in ferromagnetic and ferroelectric materials.[15]

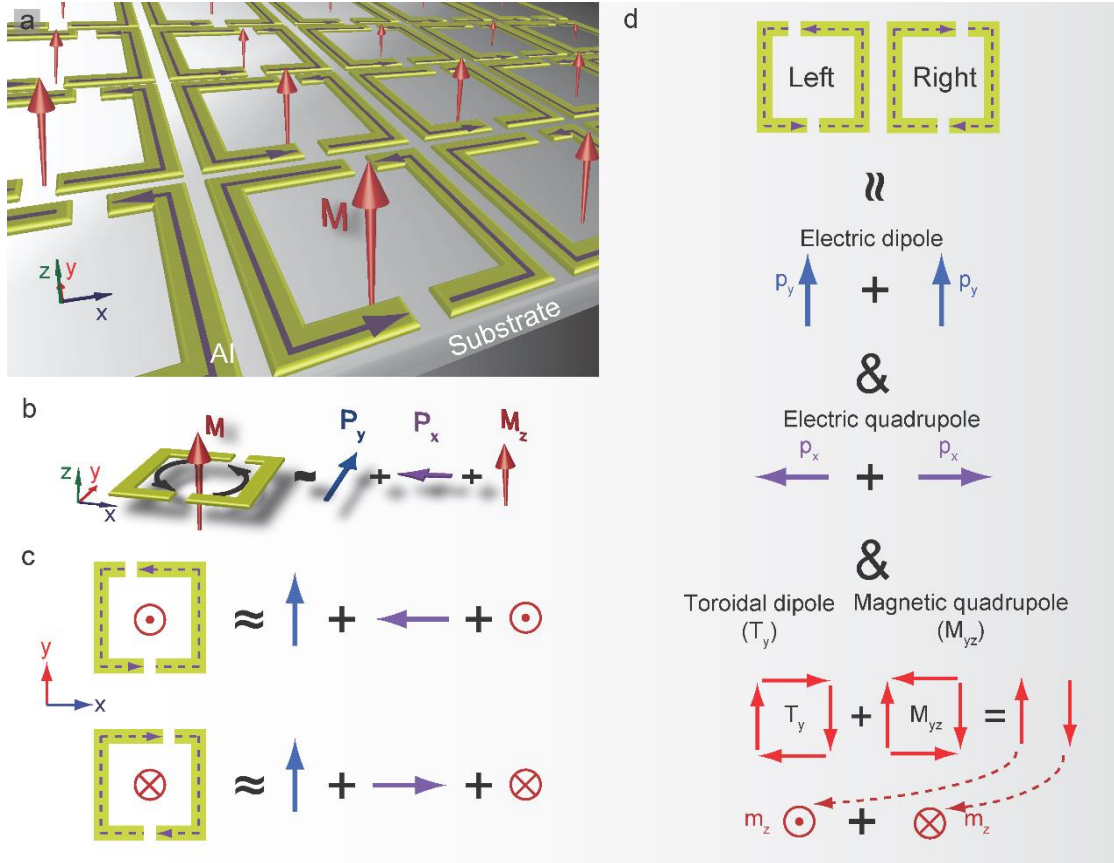

**FIG. 1. Complex multipole excitations in terahertz asymmetric split ring metamaterials at Fano resonance.** (a) Artistic impression of the TASR metamaterial with surface currents. (b) Simplified multipole representation of a single TASR. Black curved arrows denote opposing currents that oscillate in the two metallic segments of the ring. The ring can be represented as a superposition of two electric dipoles ($P_x$ and $P_y$) and a magnetic dipole ($M_z$). (c) Multipole representation of the left- and right-oriented split ring driven by vertically polarized radiation at normal incidence. The orientation of driving radiation constrains the vertical electric dipole to be the same in both cases, but the orientations of horizontal electric dipole ($P_x$) as well as magnetic dipole ($M_z$) change. (d) Multipole decomposition of a supercell consisting of left and right TASRs, illustrating how individual rings can be combined to create complex excitations accessible with normal incident plane wave.

Our analysis will focus on two kinds of TASRs, which are mirror images of each other and defined according to the positions of top split gap as '*left*' and '*right*'. In all cases the metamaterial will be driven by vertically polarized radiation (along *y*-axis in Fig. 1a) at normal incidence. Therefore, the vertical electric dipole induced in the '*left*' and '*right*' TASRs will be identical in the two kinds of resonators, but the orientation of accompanying horizontal electric dipole and out-of-plane magnetic dipole will be reversed as illustrated in Fig. 1c. The link between the geometry of the TASRs and the orientation of electric ($P_x$) and magnetic dipoles ($M_z$) induced in them allows building complex multipole excitations out of individual TASR, and investigating Ising model interactions. For example, a combination of a '*left*' and '*right*' TASR loops, as shown in Fig. 1d, corresponds to *xy*-electric quadrupole, and a superposition of toroidal dipole and magnetic quadrupole.

We begin by considering four metamaterials with composite periodic supercells ($S_1$, $S_2$, $S_3$, $S_4$) that consist of four TASRs in square lattice arrangement (left column of Fig. 2). $S_1$ consists exclusively of *right* TASRs, whereas the other three metamaterials represent all possible permutations in which two out of four TASRs in each supercell are '*right*' and the other two are '*left*'. The left column in Fig. 2 denotes the orientation of the magnetic dipoles of the four TASRs at Fano resonance (for $S_1$ all magnetic dipoles point down). Drawing parallel with magnetic materials, we note that metamaterial $S_1$ can be regarded as a representation of a ferromagnetic material, with magnetic dipoles of all domains (TASRs) pointing in the same direction, whilst the other three metamaterials ($S_2$-$S_4$) correspond to antiferromagnetic materials. Furthermore, as shown in Fig. 1b and 1c, single TASR corresponds to both horizontal electric and out-of-plane magnetic dipoles. The metamaterials $S_1$-$S_4$ are therefore metamaterial analogues of multiferroic materials.[15]

The strong interactions between the unit cells (nearest-neighbor interaction) in TASR metamaterials[16] make it virtually impossible to use the hybridization approach[17] to explain how complex TASR metamaterial response arises out of dispersion properties of individual metallic segments. Instead, we analyze the resonant response based on interaction energy. The dynamics of each charged particle in the metamaterial is governed by the Lagrangian $L = L_k - \int d^3r \left( \rho\phi - \vec{A}\cdot\vec{J} \right)$ [18], where $L_k$ is the kinetic part of the Lagrangian, $\rho$, $\vec{J}$ are charge and current density due to particles, and $\phi$, $\vec{A}$ are scalar and vector potentials. The above Lagrangian is a general modality, and we simplify it by assuming that (1) response of each individual TASR in the metamaterial, when driven near the resonant frequency, is well-approximated by a simple harmonic oscillator with resonant angular frequency $\omega_0$; (2) metamaterial is driven by plane wave at normal incidence. So the excitation of all unit cells can be described by a single dynamic variable $X = X(t)$ (ignoring the edge effects). The Lagrangian then becomes:

$$L = \frac{1}{2}\dot{X}^2 - \frac{\omega_0^2}{2}X^2 - \int_{inter-resonator} d^3r \left( \rho\phi - \vec{A}\cdot\vec{J} \right) \tag{1}$$

The first part of the Largangian ensures response of a simple harmonic oscillator with angular frequency $\omega_0$, and the second part of the Lagrangian couples the metamaterial to incident radiation as well as taking into account interactions between the constituent unit cells of the metamaterial.

As illustrated in Fig. 1, each TASR of the metamaterial, at Fano resonance can be approximated as electric and magnetic dipoles. Associating the variable $X(t)$ with induced charge density, it is possible to show (see supplementary material) that the resonant frequency of the whole metamaterial will be given by

$$\Omega = \sqrt{\frac{\omega_0^2 + \alpha}{1 - \gamma/\omega_0^2}} + \Omega_{near} \approx \omega_0 + \frac{\alpha + \gamma}{2\omega_0} + \Omega_{near} \qquad (2)$$

where $\alpha$ is related to electric dipole-dipole interaction energy and $\gamma$ is related to magnetic dipole-dipole interaction energy between the TASRs. An additional frequency shift $\Omega_{near}$ is added to account for the left-over high-order multipole interactions between the nearest-neighbor TASRs. Coming back to the Ising model, we note that $\Omega_{near}$ mimics the exchange interactions between the neighboring electrons, whilst $\frac{\alpha + \gamma}{2\omega_0}$ describes the normal long-range dipole-dipole interactions. Furthermore, the overall resonant frequency of the metamaterial becomes indicative of the strength of interactions between its constituents. One therefore obtains a correspondence between the interaction energy considered in the case of static Ising models, and the resonant frequency shift observed in dynamic TASRs and other metamaterial systems.

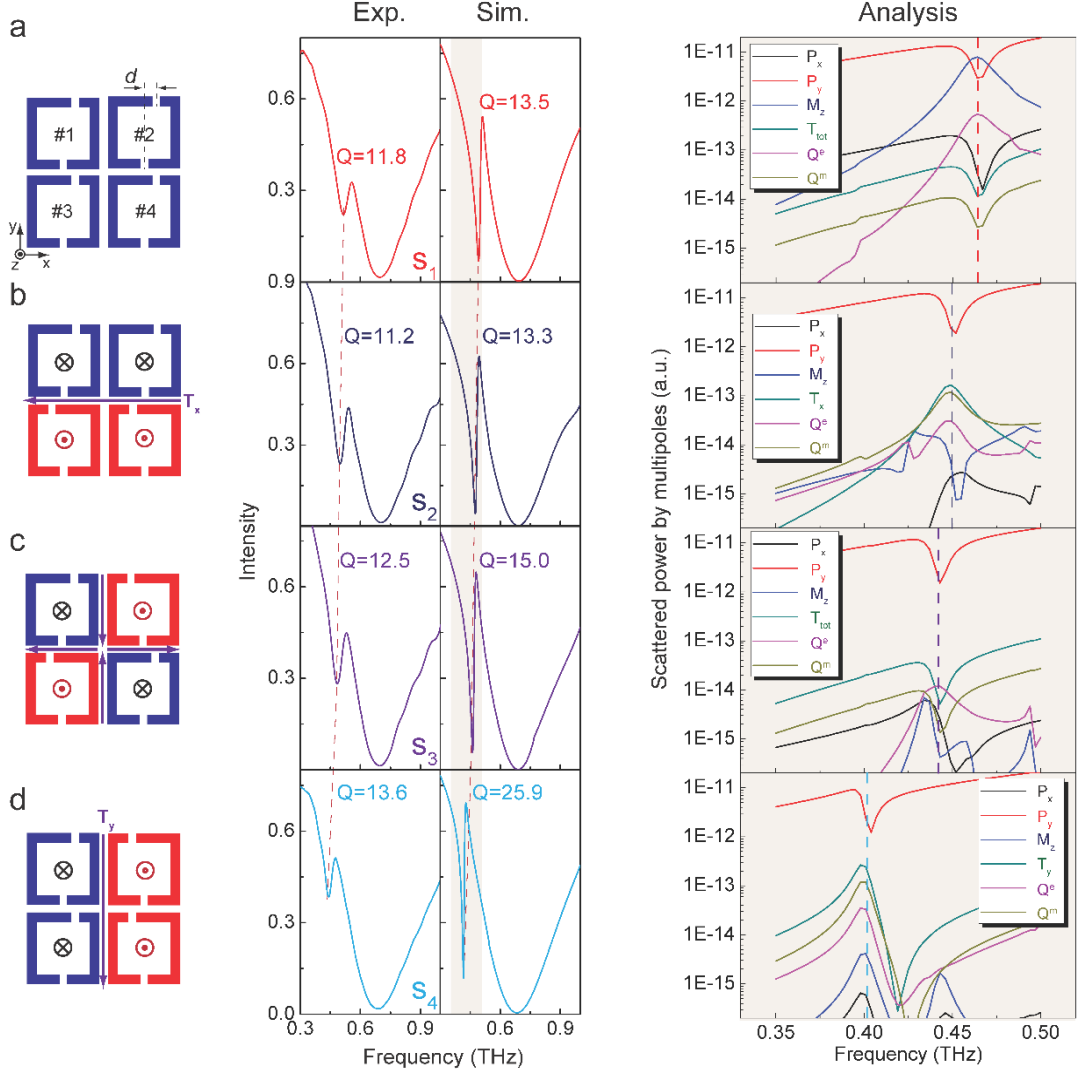

**FIG. 2. Metamaterial analogues of multiferroic materials. Left column:** (a) One supercell ($S_1$) with four identical '*right*' TASRs. The four TASRs are evenly distributed in the 160×160 μm lattice. The TASR is made of aluminum with wire width 6 μm, length 60 μm, thickness 0.2 μm and gap width 3 μm. The asymmetry is obtained by displacing the top gap with distance $d$ = 10 μm. (b-d) Supercells $S_2$-$S_4$ obtained by mirroring two out of four TASRs in $S_1$ around the vertical axis. The bull's eye (⊙) and crossed circle (⊗) pictograms in $S_2$-$S_4$ denote the orientation of magnetic dipole in the TASR loops at Fano resonance. Alongside out-of-plane magnetic dipoles, the loops will also correspond to electric dipoles as illustrated in Fig. 1. **Central column:** Measured and modelled transmission spectra of the metamaterials with annotated quality factor ($Q$) of the Fano transmission dip. **Right Column:** Multipole representation of each supercell in the vicinity of Fano mode.

In order to verify the understanding, four sets of metamaterials $S_1$-$S_4$ were designed with diverse multipole interactions, fabricated by standard photolithography, and characterized using an antenna based terahertz time-domain spectroscopy system (THz-TDS; see supplementary material). Additionally, the response of the four metamaterials was also modelled on a numerical solver (CST Microwave Studio). The measured and computed transmission intensity spectra of the four metamaterials are obtained from $T(\omega) = (\tilde{t}_s(\omega)/\tilde{t}_r(\omega))^2$ where $\tilde{t}_s(\omega)$ and $\tilde{t}_r(\omega)$ are the transmission amplitude after Fourier transform, respectively, as shown in the central column of Fig. 2. We first focus our attention on the spectral position of the metamaterial transmission dips, and it is visualized that there is a resonance frequency deviation from $S_1$ to $S_4$ for the Fano resonance in experiments as well as in simulations. Different supercell configurations enable various nearest-neighbor interaction energy which would dominate the collective resonance frequency at the Fano mode, and the corresponding resonances occur at frequencies $\Omega_1/2\pi = 0.52$ THz, $\Omega_2/2\pi = 0.50$ THz, $\Omega_3/2\pi = 0.48$ THz, and $\Omega_4/2\pi = 0.44$ THz, for metamaterials $S_1$-$S_4$ in experiments, respectively.

The four metamaterials in Fig. 2 present a perfect test-bed for the investigation of the interactions that govern the dynamics of TASR and similar metamaterials. For example, comparing it is possible to show (see supplementary material) that dipole-dipole interaction energy of the $S_2$ metamaterial (Fig. 2b) is lower than that of $S_4$ (Fig. 2d) metamaterial, yet the resonant frequency of $S_2$ is higher than that of $S_4$. Equation 2 therefore suggests that resonant frequency shift, and thus the nearest-neighbor unit cell interactions are dominated by near-field terms (related to higher order multipoles, $\Omega_{near}$) rather than by dipole-dipole interactions as is commonly assumed in case of asymmetric split ring metamaterials.[9-14] The dominant role of the strong nearest-neighbor interaction also reflects on the tailoring of far-field radiation patterns via deploying the multipoles in different metamaterial configurations.

As presented in Fig. 1d, the resultant multipoles at the Fano resonance are tailored by the strong interaction between resonators, which would enable the manipulation of radiative properties of the Fano mode. The respective multipole analysis (see supplementary materials) is performed for the four cases with distinct magnetic and electric dipole interactions as shown in the third column of Fig. 2. Consistent with Fig. 1b for $S_1$ with identical resonators, the electric dipole along $y$-axis ($P_y$) dominates the scattering, together with a magnetic dipole along $z$-axis ($M_z$). In this coherent resonance, the electric dipole has been largely suppressed which thus results in a low-loss Fano resonance characterized by the quality factor ($Q$, $Q$ = 11.8). A higher order mode, electric quadrupole ($Q^e$), appears as a result of superposition of electric dipole along $x$ axis ($P_x$).

As for $S_2$, a pair of anti-aligned magnetic dipoles, created by two co-planar current loops, corresponds to a superposition of a toroidal dipole[19-22] and a magnetic quadrupole. It has also been shown that in case of planar metamaterials the substrate can act as an additional magnetic dipole, making the combination of two co-planar currents *on* the substrate, to respond predominantly as an in-plane toroidal dipole.[23]

The two pairs of anti-aligned magnetic dipoles in supercell $S_2$ therefore add up to a net toroidal dipole along the $x$-axis ($T_x$). It is verified from the multipole analysis where the suppression of $M_z$ and enhancement of $T_x$ accompanied with $Q^M$ are visualized. Note that electric dipole scattering far exceeds the scattering rate of all the other multipoles, and the toroidal dipole is orthogonal to electric dipole, the additional scattering through toroidal dipole and magnetic quadrupole therefore have negligible effect on radiation loss, hence the $Q$ factor of $S_2$ is almost the same as that of $S_1$.

As illustrated in Fig. 2c for $S_3$, the configuration of magnetic dipoles in the supercell is such that it gives rise to two pairs of toroidal dipoles at resonance. One pair is aligned along the $y$-axis, whilst the other lies along the $x$-axis. In both pairs the two individual toroidal dipoles oscillate in anti-phase, leading to net suppression of toroidal dipole both along the $x$- and the $y$-axis. This analysis is supported by the drop in the scattered power of toroidal dipole at resonance (see analysis in Fig. 2c). The resultant experimental and simulated transmission spectra show that the $Q$ factor is slightly increased ($Q = 12.5$ in experiment). The scattered power of multipoles also confirms that all the multipoles except electric dipole are suppressed which effectively reduces the radiative losses from these higher order multipoles, and thus resulting in the slightly increased $Q$ factor of Fano resonance.

A more interesting scenario exists in $S_4$, where the toroidal dipole along $y$ axis ($T_y$) is enhanced. This toroidal dipole has the same orientation as the dominant electrical dipole so that interference will occur between them. A destructive interference ($T_y$ and $P_y$ oscillate anti-phase, see supplementary materials) effectively reduces the radiative loss, enabling a much improved $Q$ factor of the resultant Fano resonance ($Q$ = 25.9 in simulation, and $Q$ = 13.6 in experiment) as shown in Fig. 2d. Such an in-plane toroidal dipole parallel to the electric dipole could be tailored for a nonradiative anapole configuration by tuning their intensity and phase for a completely destructive interference.[24]

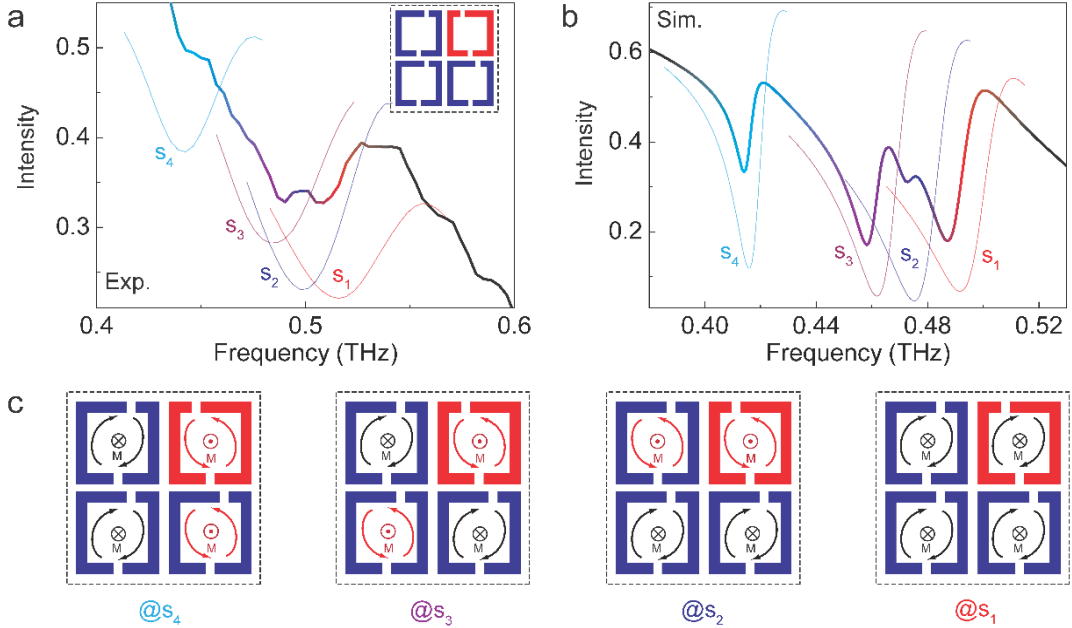

**FIG. 3. Scenario of asymmetric supercell $S_5$.** (a) Experimental transmission spectrum of $S_5$ with measured spectra of $S_1$, $S_2$, $S_3$ and $S_4$ as background for comparison. Only one resonator is mirrored in supercell $S_5$. (b) Simulated transmission intensity spectrum of $S_5$ with the background Fano resonance spectra of $S_1$, $S_2$, $S_3$ and $S_4$ for comparison. (c) Schematic current distributions for $S_5$ at the four different resonance frequencies, which would enable the similar multipole configurations as that of $S_1$ to $S_4$.

By mimicking the dynamic dipoles in metamaterial to an Ising model, we have investigated the effects of nearest-neighbor interactions on Fano resonance in terms of the resonant energy as well as the quality factor where higher order multipoles (including toroidal dipole) play a dominant role. All the four Fano resonances under the respective dipole configurations reveal distinct properties that could behave as the eigen modes of the respective scenarios. Benefitting from the design flexibility of metamaterials, a more complex scenario is probed under the Ising model analogue, where we observe a hyperfine splitting Fano mode in absence of external static magnetic/electric field analogous to the Zeeman/Stark effects. As the inset image shown in Fig. 3a for supercell $S_5$ with only one resonator mirrored, the measured transmission spectrum shows four separated dips that correspond to the eigen Fano modes supported by supercells $S_1$, $S_2$, $S_3$ and $S_4$, respectively. Such a hyperfine splitting Fano spectrum is clearly captured in simulations as shown in Fig. 3b, where the resonance dips coincide with the respective eigen Fano modes of supercells $S_1$, $S_2$, $S_3$ and $S_4$. We note that the resonance footprints in experiments are weaker than that in the simulated spectrum due to the finite unit cells (assuming infinite in simulations) under illumination in experiments and imperfection of fabricated samples (see supplementary materials).[14,25]

The dominant role of nearest-neighbor interaction between unit cells is clearly revealed in the asymmetric supercell $S_5$ where all the possible interaction channels in a 2×2 supercell are activated simultaneously. Different electric and magnetic dipole distributions are resumed at the four resonance frequencies as indicated in Fig. 3c, that correspond to various multipole configurations. In addition to the first part of the Largangian that originates from the individual harmonic oscillator ($\omega_0$), nearest-neighbor interaction under the Ising model analogue provides an extra freedom to engineer the mode properties through manipulating the spatial topology, and more promising phenomena could be predicted for higher order supercell arrangement as well as three-dimensional configuration.

Due to the low density of the "*left*" resonator in $S_5$, the coupling strength of the four Fano modes to free space is relatively weak,[25,26] leading to the difficulty in capturing the spectral features in experiments. One solution is by tailoring the gap displacement (*d*) that is readily accessed in the artificially designed metamaterial. The larger displacement of the gap would give rise to a stronger coupling of electric dipole to free space in the collectively oscillating mode, and thus result in a more pronounced spectral feature (see supplementary material).[27] The flexibility of metamaterials also reflects on the scalability of operation bands into infrared and visible regimes via tailoring the geometric size of unit cells, where the hyperfine Fano splitting could be applied for many valuable applications in biosensing,[28] imaging,[29] and selective thermal emitter.[30] Since Fano resonance has been demonstrated as an excellent platform for refractometric sensing,[28,31] we summarized the performance of the four Fano modes in terms of sensitivity in Table 1. We note that each mode reveals exclusive sensitivity and spectral quality factor mediated by various multipoles, which provides multispectral fingerprints and resolutions.

Table 1. Performance of the four Fano modes

|  | S$_1$ | S$_2$ | S$_3$ | S$_4$ |
|---|---|---|---|---|
| **Frequency (THz)** | 0.49 | 0.47 | 0.46 | 0.42 |
| **Sensitivity$^a$ (GHz/RIU)** | 45.5 | 41.3 | 39.6 | 37.5 |
| **$Q$ factor** | 13.5 | 13.3 | 15.0 | 25.9 |

$^a$Sensitivity is for refractometric sensing with 20 μm thick analyte on top of metasurface.

In conclusion, we have investigated a metamaterial analogy of static Ising model in terms of Fano resonance. Strong nearest-neighbor interactions in a 2D metamaterial array dominate the collective resonance energy as well as mode quality factors. A hyperfine splitting Fano spectrum is observed in the absence of external magnetic/electric field due to simultaneous activation of all the interaction configurations, that provides a flexible path for multispectral excitation and broadband applications. A finer splitting Fano mode could be expected at a higher order supercell (3×3, 4×4, *et al*), which would enable an excellent platform for ultrasensitive sensing, super-resolution imaging, and multiband selective thermal emission. This work also demonstrates a potential path to investigate the microscopic static interactions between particles by using the macroscopic artificial electromagnetic resonators.

**Acknowledgements**

The authors acknowledge research funding support from Singapore MOE Grants No. MOE2015-T2-2-103 and UK's Engineering and Physical Sciences Research Council (Grant Nos. EP /G060363/1 and EP /M009122/1).

(2014).

[27]  L. Cong, M. Manjappa, N. Xu, I. Al-Naib, W. Zhang, and R. Singh, Advanced Optical Materials **3**, 1537 (2015).

[28]  C. Wu, A. B. Khanikaev, R. Adato, N. Arju, A. A. Yanik, H. Altug, and G. Shvets, Nature materials **11**, 69 (2012).

[29]  Y. Hiraoka, T. Shimi, and T. Haraguchi, Cell Structure and Function **27**, 367 (2002).

[30]  X. Liu, T. Tyler, T. Starr, A. F. Starr, N. M. Jokerst, and W. J. Padilla, Phys. Rev. Lett. **107**, 045901 (2011).

[31]  R. Singh, W. Cao, I. Al-Naib, L. Cong, W. Withayachumnankul, and W. Zhang, Appl. Phys. Lett. **105**, 171101 (2014).